\def\e{{\rm e}}
\def\del{\partial}
\def\half{{1\over2}}
\def\cosb{\cos\beta}
\def\sinb{\sin\beta}

\def\expecv#1{\langle #1 \rangle}
\def\absv#1{\left|#1\right|}

\def\llt{\mathrel{<\kern-0.5em <}}   
\def\ggt{\mathrel{>\kern-0.5em >}}   
\def\gtsim{\mathrel{\hbox{\raise0.2ex
\hbox{$>$}\kern-0.75em\raise-0.9ex\hbox{$\sim$}}}}
\def\ltsim{\mathrel{\hbox{\raise0.2ex
\hbox{$<$}\kern-0.75em\raise-0.9ex\hbox{$\sim$}}}}
\newcommand{\lw}[1]{\smash{\lower2.ex\hbox{#1}}}
\newcommand{\PRD}[3]{Phys. Rev. {\bf D{#1}} (19{#2}) {#3}}

\newcommand{\NPB}[3]{Nucl. Phys. {\bf B{#1}} (19{#2}) {#3}}
\newcommand{\PLB}[3]{Phys. Lett. {\bf B{#1}} (19{#2}) {#3}}
\newcommand{\PTP}[3]{Prog. Theor. Phys. {\bf {#1}} (19{#2}) {#3}}

\DeclareMathAlphabet{\mib}{OML}{cmm} {b}{it}
\documentstyle[epsf,12pt]{article}
\makeatletter

\@addtoreset{equation}{section}
\makeatother
\begin{document}
\begin{titlepage}
\begin{flushright}
SAGA--HE--121\\
\date={April 19,1997}
\end{flushright}
\vspace{24pt}
\centerline{\Large\bf CP Violating Bubble Wall and Electroweak Baryogenesis}
\vspace{24pt}
\begin{center}
{\bf Koichi~Funakubo$^{a,}$\footnote{e-mail: funakubo@cc.saga-u.ac.jp},
 Akira~Kakuto$^{b,}$\footnote{e-mail: kakuto@fuk.kindai.ac.jp},
 Shoichiro~Otsuki$^{b,}$\footnote{e-mail: otks1scp@mbox.nc.kyushu-u.ac.jp}\\
 and Fumihiko~Toyoda$^{b,}$\footnote{e-mail: ftoyoda@fuk.kindai.ac.jp}}
\end{center}
\vskip 0.8 cm
\begin{center}
{\it $^{a)}$Department of Physics, Saga University,
Saga 840 Japan}
\vskip 0.2 cm
{\it $^{b)}$School of Engineering, Kinki University in Kyushu,
Iizuka 820 Japan}
\end{center}
\baselineskip=20pt
\vskip 1.0 cm
\bigskip\bigskip
\abstract{
The electroweak baryogenesis depends on the profile of the $CP$-violating
bubble wall created at the first order phase transition. We attempt to
determine it by solving the coupled equations of motion for the moduli 
and phases of the two Higgs doublets at the transition temperature.
A variety of $CP$-violating bubble walls are classified by boundary
conditions. We point out that a sufficiently small explicit
$CP$ violation gives nonperturbative effects to yield the baryon
asymmetry of the universe.}
\vfill\eject
\end{titlepage}
\baselineskip=18pt
\setcounter{page}{2}
\setcounter{footnote}{0}
\section{Introduction}
To generate the baryon asymmetry of the universe (BAU) starting from a
baryon-\break
symmetric universe, Sakharov's three conditions must be met.
If there was no $B-L$ down to the electroweak era, baryogenesis mechanism
should work at the electroweak phase transition (EWPT) for the present
universe to be realized\cite{reviewEB}. Even if some $B-L$ existed, 
what happened at the EWPT would affect the present BAU. 
Among the conditions, to have sufficient $CP$ violation necessary for
the BAU will require
some extension of the Higgs sector in the minimal standard model.
Further, the minimal standard model satisfying the present lower bound
on the Higgs mass cannot make the EWPT first order and cannot suppress
the sphaleron process after that.
Two-Higgs-doublet extension of the standard model is the simplest
extension, which can yield the Higgs-sector $CP$ violation,
and includes the Higgs sector of the minimal supersymmetric standard
model (MSSM) as a special case.
Here we concentrate on such two-Higgs-doublet models.\par
At the first-order EWPT, a bubble wall dividing the broken and 
symmetric phases is created and grows to convert the whole universe
into the broken phase. The profile of the bubble wall will be
characterized by the classical configuration of the Higgs and
gauge fields.
It is essential in any scenario of the electroweak baryogenesis 
to know the spacetime-varying profile of the bubble wall, in which
the relative phase of the Higgs scalars violates $CP$.
In literatures, some functional forms of the profile were
assumed\cite{NKC}\cite{FKOTTa}.
They should be determined, however, by the dynamics of the gauge-Higgs
system. That is, the bubble wall will be a solution to the
classical equations of motion which has the effective potential as
the potential for the Higgs sector.
The observable $CP$ violations in our world offer the boundary
condition in the broken phase for these equations of motion.\par
In a previous paper\cite{FKOTTb}, assuming some appropriate form
of the effective potential, we found a solution such that $CP$-violating
relative phase becomes as large as $O(1)$ around the wall while it
completely vanishes in the broken and symmetric phase limits.
In the solution, the moduli were fixed to be the kink shape. Such a 
solution, which we call as a ``solution with kink ansatz'', is
interesting since it does yield sufficient amount of the chiral charge
flux through the wall surface,
which will be turned into the baryon number in the symmetric phase
region by the sphaleron transition.
The solution emerges when a mechanism similar to the spontaneous $CP$
violation operates in the intermediate region.
For such a mechanism to work, some of the parameters in the effective
potential are restricted similarly to the case of the spontaneous $CP$
violation at zero temperature, although less constrained at finite
temperature.\par
The next task will be to determine the moduli themselves of the
Higgs fields dynamically as well as the relative phase of them.
In this paper, we solve the coupled equations of motion for the moduli
and the relative phase of the two Higgs fields
by imposing a discrete symmetry between the two Higgs doublets to reduce
the dynamical degrees of freedom.
With the same parameters of the effective potential as those with
the kink ansatz, we present a solution, which we call as a
``solution without kink ansatz'', having larger $CP$-violating phase
and non-kink type moduli. As expected, this solution has lower energy
than that with the kink ansatz.\par
We also give another class of new solutions, which are caused by a
similar mechanism of $CP$ violation in the intermediate region as above
but are free from the constraints of zero-temperature spontaneous $CP$
violation. The nontrivial phase dependence of cubic terms
in the moduli, which is essential to the first-order EWPT, plays a
crucial role in this mechanism.
Such terms will be induced from bosonic loops, especially those of
the scalar partners of quarks and leptons in the MSSM and the extra
Higgs scalars.\par
If $CP$ violation is completely spontaneous, a $CP$-violating bubble
is created with the same probability as its $CP$-conjugate one.
Then the net baryon number would result in zero. We show that
a tiny explicit $CP$ violation, which is consistent with the
present bound on the neutron electric dipole moment (EDM), can
nonperturbatively resolve the degeneracy between the $CP$-conjugate
pair of the bubbles to leave a sufficient BAU after the EWPT.\par
In section 2, we postulate the effective potential of the Higgs sector
and present the equations of motion. 
In section 3, we classify possible solutions according to the
boundary conditions, and give some examples of
``solutions without kink ansatz''.
We also investigate solutions in the presence of explicit $CP$ breaking,
and point out the importance of the breaking to generate the BAU.
Section 4 is devoted to conclusion and discussions. 
\section{The Model}
One of the most attractive features of electroweak baryogenesis is
the fact that it depends only on physics which can be tested by
experiments at present as well as in the near future.
We emphasize that it is $CP$ violation at the
first-order EWPT, not that observed in laboratories, which 
affects the generated baryon asymmetry. 
The latter is restricted by various observables such as the neutron
EDM, while the former is not {\it directly}.
Here we attempt to relate the observable $CP$ violation to that
at the EWPT.\par
\subsection{Ansatz for the effective potential}
Given a model, one can calculate the effective potential near the
phase transition temperature, which determines the order of the EWPT
as well as its dynamics.
We expect that the bubble wall will be well approximated by a
solution to the classical equations of motion derived from the effective
Higgs potential, which consists of the tree-level potential with
radiative corrections and finite-temperature contributions.
Even if the tree-level Higgs potential is $CP$ conserving,
the corrected potential parameters may predict spontaneous
$CP$ violation, or the corrections may bring effects of various explicit
$CP$-violating parameters, such as those in the soft
supersymmetry breaking parameters, into the effective potential.
The $CP$ violation in the Kobayashi-Maskawa matrix could also induce
$CP$-violating term in the effective potential through higher order
diagrams.\par
Here we postulate a functional form of the effective potential, which
is equipped with the feature of first-order EWPT.
We assume that it is well approximated by a gauge-invariant polynomial
up to the fourth order of the moduli of the Higgs fields.
The most general gauge-invariant Higgs potential is
\begin{eqnarray}
 V_0&=&
 m_1^2\Phi_1^\dagger\Phi_1 +  m_2^2\Phi_2^\dagger\Phi_2
 +( m_3^2\Phi_1^\dagger\Phi_2 + {\rm h.c.} )                  \nonumber \\
 &+&
 \half\lambda_1(\Phi_1^\dagger\Phi_1)^2+\half\lambda_2(\Phi_2^\dagger\Phi_2)^2
 +\lambda_3(\Phi_1^\dagger\Phi_1)(\Phi_2^\dagger\Phi_2) \label{eq:tree-V}\\
 &-&
 \lambda_4(\Phi_1^\dagger\Phi_2)(\Phi_2^\dagger\Phi_1)
 -\left\{ \half\lambda_5(\Phi_1^\dagger\Phi_2)^2
 +\bigl[\lambda_6(\Phi_1^\dagger\Phi_1)+\lambda_7(\Phi_2^\dagger\Phi_2)\bigr]
  (\Phi_1^\dagger\Phi_2) + {\rm h.c.} \right\},                 \nonumber
\end{eqnarray}
where $m_1^2, m_2^2, \lambda_1, \lambda_2, \lambda_3, \lambda_4 \in {\bf R}$
and $m_3^2, \lambda_5, \lambda_6, \lambda_7 \in {\bf C}$, 
three of their phases are independent and yield the explicit $CP$ violation.
To avoid the tree-level Higgs-mediated flavor-changing neutral current
(FCNC) interactions naturally, one may impose the discrete
symmetry on the lagrangian\cite{GlashowWeinberg}, which is broken by
$\lambda_{6,7}\not=0$ and softly by $m_3^2\not=0$.
The Yukawa interactions should have that symmetry not to induce 
divergent $\lambda_{6,7}$-terms.
In the MSSM, the parameters in the Higgs potential are given by
\begin{eqnarray}
 & &\lambda_1=\lambda_2={1\over 4}(g^2+{g^\prime}^2),\qquad
    \lambda_3={1\over4}(g^2-{g^\prime}^2),\qquad
    \lambda_4={1\over2}g^2,   \nonumber\\
 & &\lambda_5=\lambda_6=\lambda_7=0,   \label{eq:MSSM-V0}
\end{eqnarray}
where $g$ ($g^\prime$) is the $SU(2)$ ($U(1)$) gauge coupling,
the mass parameters are completely arbitrary, and the Yukawa
interactions respect the discrete symmetry.
Hence $CP$ is not violated by the Higgs sector at the tree level in 
the MSSM.
Nonzero $m_3^2$ will induce finite corrections to $\lambda_{5,6,7}$,
which violates $CP$ explicitly or spontaneously.\par
At finite temperature, there arise new kind of corrections, which 
lead to the first-order EWPT. In general, the one-loop 
finite-temperature corrections to the effective potential have a form
\begin{equation}
 {\bar V}({\mib v};T) =
 {{T^4}\over{2\pi^2}}\sum_A d_A I_A(a_A^2),  \label{eq:finiteTcorr}
\end{equation}
where $A$ stands for particle species running through the loop,
\begin{equation}
 I_{B,F}(a^2) \equiv
  \int_0^\infty dx\, x^2\log\left(1\mp{\e}^{-\sqrt{x^2+a^2}}\right)
     \label{eq:def-Is}
\end{equation}
with $a_A=m_A({\mib v})/T$, ${\mib v}$ being the VEV of the Higgs
fields, and $d_A$ counts the degrees of freedom with
sign (e.g., $d_W=6$, $d_Z=3$ and $d_t=-6$). 
One can expand $I_{B,F}(a^2)$ around $a^2=0$ to observe that the 
expansion of $I_B$ contains $(a^2)^{3/2}$-term, which originates
from the Matsubara zero mode of the loop momentum\cite{DJ}.
Hence if $m^2({\mib v})\sim 0$ at ${\mib v}\sim0$, the effective 
potential has ${\mib v}^3$ terms with the correct sign, which is 
the characteristic of the first-order phase transition.
Here we assume that the effective potential at the EWPT is approximated 
by the polynomial of the form of (\ref{eq:tree-V}) plus ${\mib v}^3$-terms
irrespective of the validity of the perturbation and the high-temperature
expansion.\par
First of all, we shall concentrate on the case with no explicit $CP$
violation in the effective potential. Then the ansatz for it is
\begin{eqnarray}
 V_{\rm eff}(\rho_i,\theta)&=&
 \half m_1^2\rho_1^2+\half m_2^2\rho_2^2
 + m_3^2\rho_1\rho_2\cos\theta          \nonumber\\
 & &
 +{{\lambda_1}\over8}\rho_1^4+{{\lambda_2}\over8}\rho_2^4 +
 +{{\lambda_3-\lambda_4-\lambda_5\cos(2\theta)}\over4}\rho_1^2\rho_2^2
              \nonumber\\
 & &
 -\half(\lambda_6\rho_1^3\rho_2+\lambda_7\rho_1\rho_2^3)\cos\theta \nonumber\\
 & &
 -\left[A\rho_1^3+(B_0+B_1\cos\theta+B_2\cos(2\theta))\rho_1^2\rho_2
       \right.           \nonumber\\
 & &\left.\qquad\qquad
 +(C_0+C_1\cos\theta+C_2\cos(2\theta))\rho_1\rho_2^2+D\rho_2^3\right],
       \label{eq:ansatz-Veff}
\end{eqnarray}
where we parameterize the Higgs VEV as\footnote{Here we assume
that $U(1)_{\rm em}$ gauge symmetry is unbroken at the EWPT.}
\begin{equation}
 \expecv{\Phi_i}=
 \pmatrix{0 \cr {1\over{\sqrt{2}}}\rho_i{ e}^{i\theta_i}\cr},\qquad
 \theta = \theta_1-\theta_2.        \label{eq:def-VEV0}
\end{equation}
As we noted, the parameters in (\ref{eq:ansatz-Veff}) should be 
understood to contain radiative and finite-temperature corrections,
which can be calculable once the model is specified.
According to the high-temperature expansion, $B_a$ and $C_a$ may arise
from the bosonic loops, among which $\theta$-dependent terms, $B_{1,2}$ and
$C_{1,2}$, could arise only from those of squarks, sleptons and 
Higgs scalars, that mix scalars and pseudoscalars in the MSSM.
Viewed as a function of $\cos\theta$, the effective potential is 
written as
\begin{eqnarray}
   V_{\rm eff}(\rho_i,\theta) &=&
 -\left[{{\lambda_5}\over2}\rho_1^2\rho_2^2 +
        2(B_2\rho_1^2\rho_2+C_2\rho_1\rho_2^2)\right]  \nonumber\\
  & &\times
  \left\{\cos\theta
  +{{-2m_3^2+\lambda_6\rho_1^2+\lambda_7\rho_2^2+2(B_1\rho_1+C_1\rho_2)}
    \over{2\lambda_5\rho_1\rho_2+8(B_2\rho_1+C_2\rho_2)}}\right\}^2\nonumber\\
  & &+ \cdots.     \label{eq:Veff-costheta}
\end{eqnarray}
When
\begin{equation}
  \lambda_5\rho_1^2\rho_2^2 + 4(B_2\rho_1^2\rho_2+C_2\rho_1\rho_2^2)
  <0,        \label{eq:cond-CPv1}
\end{equation}
and
\begin{equation}
  \absv{{-2m_3^2+\lambda_6\rho_1^2+\lambda_7\rho_2^2+2(B_1\rho_1+C_1\rho_2)}
    \over{2\lambda_5\rho_1\rho_2+8(B_2\rho_1+C_2\rho_2)}} < 1,
               \label{eq:cond-CPv2}
\end{equation}
$CP$ is spontaneously broken.
At zero temperature, $B_a=C_a=0$ and
$(\rho_1,\rho_2)=(v_0\cos\beta_0,$
$v_0\sin\beta_0)$ so that these conditions are
reduced to the well-known ones,
\begin{equation}
 \lambda_5 < 0,\quad\mbox{and}\quad
 \absv{{-2m_3^2/v_0^2+\lambda_6\cos^2\beta_0+\lambda_7\sin^2\beta_0}\over
       {2\lambda_5\sin\beta_0\cos\beta_0}}<1.  \label{eq:spontCPatzeroT}
\end{equation}
Here all the parameters are evaluated at zero temperature.
The former condition ($\lambda_5 < 0$) is hardly fulfilled, as long as
$\lambda_5\ge 0$ at the tree level, since almost all radiative corrections
except for those by the gauginos and Higgsinos contribute to make
$\lambda_5$ positive in the MSSM\cite{Maekawa}.
Even if these conditions are satisfied, such a model will be bothered
by a light scalar, which is inevitable in the presence of spontaneous
violation of a discrete symmetry\cite{GeorgiPais}. 
In contrast, the conditions (\ref{eq:cond-CPv1}) and (\ref{eq:cond-CPv2})
can be satisfied in the intermediate range, since $\rho_1$ ($\rho_2$)
varies from $v\cos\beta$ ($v\sin\beta$) to zero as going from the
broken phase to the symmetric phase region.
Since the finite-temperature corrections both of bosons and fermions
to $\lambda_5$ are positive at the one-loop level, 
requiring (\ref{eq:cond-CPv1}) and (\ref{eq:cond-CPv2}) to be satisfied
only near the bubble wall and not to be satisfied in the broken phase
would constrain some parameters in the model.
Especially in the MSSM, for the finite-temperature corrections to
negative $\lambda_5$ to be small,
some soft-supersymmetry-breaking mass parameters will be bounded from 
below\cite{FKOTc}.\par
If $CP$ is violated spontaneously, there are two kinds of bubbles
$(\rho_i,\theta)$ and $(\rho_i,-\theta)$ which are $CP$ conjugate
with each other and have the same energy density.
Such bubbles would be nucleated with the same probability, so that
no net baryon number would be generated.
An explicit $CP$ violation will resolve this degeneracy and leave
the finite BAU.
Although various types of explicit $CP$ violation in the effective
potential are possible, we restrict to a simple case that only the 
$m_3^2$ is complex and the other parameters are real.
Hence we replace $m_3^2\rho_1\rho_2\cos\theta$ in $V_{\rm eff}$
of (\ref{eq:ansatz-Veff}) by
\begin{equation}
  m_3^2\rho_1\rho_2\cos(\theta+\delta),
  \qquad (m_3^2\in{\bf R}).     \label{eq:explicitCP}
\end{equation}
In the MSSM, nonzero $\delta$ is induced from the $CP$ violation
in the scalar trilinear terms and $\absv{\delta}<10^{-2}$ does not
contradict with the present bound on the neutron EDM\cite{CPVrev}.
The effect of $\delta$ on the baryon asymmetry was discussed
in \cite{Comelli} perturbatively.
We shall see later that $\delta$ nonperturbatively discriminates
the $CP$-conjugate pair of the bubbles for some types of solutions.\par
\subsection{Equations of motion}
We suppose that the dynamics of the bubble wall is governed by
the classical equations of motion derived from the lagrangian
\begin{equation}
 {\cal L} = -{1\over4}F^a_{\mu\nu}F^{a\,\mu\nu}-{1\over4}B_{\mu\nu}B^{\mu\nu}
            +\sum_{i=1,2}\left(D_\mu\Phi_i\right)^\dagger D^\mu\Phi_i
            -V_{eff}(\Phi_1,\Phi_2;T_C),
	\label{eq:lagrangian}
\end{equation}
where,
$$
  D_\mu\Phi_i(x) \equiv (\del_\mu-ig{{\tau^a}\over2}A^a_\mu(x)
                       -i{{g^\prime}\over2}B_\mu(x))\Phi_i(x).
$$
If the bubble is spherically symmetric and sufficiently macroscopic,
it is approximated as a planar object so that the system is reduced
to one-dimensional. Further, when the bubble grows with a constant
velocity keeping the shape of the critical bubble, it can be viewed as
a static object.
For this effective one-dimensional system, the gauge fields may be
written in pure gauge forms and a solution in this case will
have lower energy than solutions with nontrivial gauge configuration.
Hence we assume that the bubble wall consists of only the Higgs 
scalar, and work in the gauge where the gauge fields are eliminated.
Since the fermions couple to gauge-noninvariant phases of the Higgs
instead of the invariant $\theta$, the $CP$ violation in the Yukawa
interactions is made unambiguous once the gauge is fixed in this 
manner.\par
Then the bubble wall profile is characterized by the space-varying
order parameters $(\rho_i(z),\theta_i(z))$, which satisfy 
the following equations of motion,
\begin{eqnarray}
 {{d^2\rho_i(z)}\over{dz^2}}-\rho_i(z)\left({{d\theta_i(z)}\over{dz}}\right)^2
  -{{\del V_{eff}}\over{\del\rho_i}} &=& 0,
        \label{eq:rho-z}  \\
 {d\over{dz}}\left(\rho_i^2(z){{d\theta_i(z)}\over{dz}}\right)
  -{{\del V_{eff}}\over{\del\theta_i}} &=& 0,
        \label{eq:theta-z}
\end{eqnarray}                                      
where $z$ is the coordinate perpendicular to the wall.
{}From the requirement that the gauge fields are pure-gauge type,
a constraint equation, ``sourcelessness condition", should be fulfilled:
\begin{equation}
  \rho_1^2(z){{d\theta_1(z)}\over{dz}}+\rho_2^2(z){{d\theta_2(z)}\over{dz}}=0.
        \label{eq:soureless-z}
\end{equation}
Now our task is to solve the equations (\ref{eq:rho-z}) and 
(\ref{eq:theta-z}) with the constraint (\ref{eq:soureless-z}) 
satisfying certain boundary conditions in the broken phase and symmetric 
phase regions.\par
Before solving the equations, we reduce the dynamical degrees of 
freedom by adopting some ansatz.
First, we require that the moduli of the Higgs scalars take kink 
shapes of the common width when $\theta$ is sufficiently small. 
In practice, we assume that when $\delta = 0$, 
\pagebreak
\begin{eqnarray}
 \theta(z) &=& 0, \nonumber \\
 \rho_1(z) &=& v\cos\beta{{1+\tanh(az)}\over2}, \label{eq:kink-rho-z} \\
 \rho_2(z) &=& v\sin\beta{{1+\tanh(az)}\over2} \nonumber
\end{eqnarray}
are solutions to (\ref{eq:rho-z}),
where $1/a$ is the common wall thickness.
This restricts some of the parameters in the effective potential
(\ref{eq:ansatz-Veff})\cite{FKOTTb}. Then the equation of motion for 
$\theta(z)$ in the background of the kink $\rho_i(z)$ is reduced to
\begin{eqnarray}
 & & 
 y^2(1-y)^2{{d^2\theta(y)}\over{dy^2}}+y(1-y)(1-4y){{d\theta(y)}\over{dy}}
          \nonumber\\
 &=&
 b\sin(\theta(y)+\delta)+[c(1-y)^2-e(1-y)]\sin\theta(y) \nonumber\\
 & &
  + [{d\over2}(1-y)^2-2f(1-y)]\sin(2\theta(y)),    \label{eq:eq-theta}
\end{eqnarray}
where we introduced a dimensionless finite-range parameter $y$ by
\begin{equation}
 y = {1\over2}\left(1-\tanh(az)\right),    \label{eq:z-y}
\end{equation}
and the parameters in (\ref{eq:eq-theta}) are defined by
\begin{eqnarray}
  b&\equiv& -{{m_3^2}\over{4a^2\sinb\cosb}},   \nonumber\\
  c&\equiv& {{v^2}\over{32a^2}}(\lambda_1\cot^2\beta+\lambda_2\tan^2\beta
            +2(\bar\lambda_3-\lambda_5)) - {1\over{2\sin^2\beta\cos^2\beta}}
      \nonumber \\
   &=&  {{v^2}\over{8a^2}}(\lambda_6\cot\beta + \lambda_7\tan\beta),
                  \label{eq:def-bcd}\\
  d&\equiv& {{\lambda_5 v^2}\over{4a^2}},   \nonumber   \\
  e&\equiv& -{v\over{4a^2}}\left({B_1\over\sinb}+{C_1\over\cosb}\right),
      \nonumber   \\
  f&\equiv& -{v\over{4a^2}}\left({B_2\over\sinb}+{C_2\over\cosb}\right),
      \nonumber   \\
  e+f&=& {v\over{4a^2\sin^2\beta\cos^2\beta}}
             ( A\cos^3\beta+B_0\cos^2\beta\sinb \nonumber\\
     &+& C_0\cosb\sin^2\beta+
             D\sin^3\beta-{{4a^2}\over v})      \nonumber
\end{eqnarray}
with $\bar\lambda_3\equiv\lambda_3-\lambda_4$.
Here $\theta(z)$ is related to $\theta_{1,2}(z)$ by
\begin{equation}
 \theta_1(z)/\sin^2\beta = -\theta_2(z)/\cos^2\beta = \theta(z)
	\label{eq:theta12},
\end{equation}
which means that $\theta_1(z)$ and $\theta_2(z)$ satisfy the
condition (\ref{eq:soureless-z}) when
$\rho_i(z)$ is the kink shape (\ref{eq:kink-rho-z}).
We refer the solution to (\ref{eq:eq-theta}) as ``the solution with
kink ansatz''. We found several solutions for various boundary 
conditions when $f=0$ in \cite{FKOTTb}.
Among them, when $d<0$ ($\lambda_5<0$), there is an interesting 
solution which connects the $CP$-conserving vacua and largely violates $CP$ in 
the intermediate range around $\rho_i(y)\sim 1/2$ by satisfying
(\ref{eq:cond-CPv1}) and (\ref{eq:cond-CPv2}). The solution has so 
large a phase as $\theta(y=1/2)\gtsim0.3$ to generate sufficient
chiral charge flux. For the same set of
parameters, we also found that a tiny explicit $CP$ violation of the 
form (\ref{eq:explicitCP}) can nonperturbatively enhance the 
lower-energy bubble over its $CP$-conjugate partner to leave 
sufficient BAU\cite{FKOTeCP}.\par
The large $\theta$ means, however, that the kink ansatz may no longer be
adequate. Thus we have to solve coupled equations for the moduli and
the phases. In order to simplify the analysis,
we impose a discrete symmetry on the effective potential under
\begin{equation}
 \rho_1\leftrightarrow\rho_2,\qquad
 \theta_1\leftrightarrow-\theta_2,     \label{eq:def-discrete-sym}
\end{equation}
and assume that it is not spontaneously broken to reduce the dynamical
degrees of freedom.
This means that the parameters in (\ref{eq:ansatz-Veff}) should satisfy
\[
 m_1^2=m_2^2\equiv m^2,\quad
 \lambda_1=\lambda_2,\quad \lambda_6=\lambda_7,
\]
\[
 A=D,\quad B_0=C_0,\quad B_1=C_1,\quad B_2=C_2, 
\]
and that $\tan\beta=1$ by definition. For
\begin{equation}
 \rho_1(z)=\rho_2(z)\equiv\rho(z)/{\sqrt{2}},\quad
 \theta_1(z)=-\theta_2(z)\equiv\theta(z)/2,
\end{equation}
we have two coupled equations:
\begin{eqnarray}
 {{d^2\rho(z)}\over{dz^2}}
 -{1\over4}\rho(z)\left({{d\theta(z)}\over{dz}}\right)^2
  -{{\del V_{eff}}\over{\del\rho}} &=& 0,     \label{eq:sym-rho-z}\\
 {1\over4}{d\over{dz}}\left(\rho^2(z){{d\theta(z)}\over{dz}}\right)
  -{{\del V_{eff}}\over{\del\theta}} &=& 0,   \label{eq:sym-theta-z}
\end{eqnarray}
while the ``sourcelessness condition" is automatically satisfied.
Here the effective potential is written in terms of the reduced numbers
of the parameters as
\begin{eqnarray}
 V_{\rm eff}(\rho,\theta)&=&
 \half\left(m^2+m_3^2\cos(\theta+\delta)\right)\rho^2
 +{1\over{16}}\left[\tilde\lambda-\lambda_5\cos(2\theta)
  -4\lambda_6\cos\theta\right]\rho^4       \nonumber\\
 & &\quad
 -{1\over{\sqrt2}}\left[A+B_0+B_1\cos\theta+B_2\cos(2\theta)\right]
    \rho^3,    \label{eq:dis-sym-Veff}
\end{eqnarray}
where $\tilde\lambda\equiv \lambda_1+\lambda_3-\lambda_4$.
The parameters in (\ref{eq:dis-sym-Veff}) will be further constrained
by requiring that $V_{\rm eff}$ should have two degenerate minima 
corresponding to the broken and symmetric phases.
We relate the parameters to those in the potential with the kink ansatz,
$(b,c,d,e,f)$, to compare ``the solutions without kink ansatz'' to
those with it. That is, some of the parameters in (\ref{eq:dis-sym-Veff})
are related to $(b,c,d,e,f)$ directly by
\begin{eqnarray}
 m_3^2 &=& -2a^2b,                   \nonumber\\
 \tilde\lambda&=&\lambda_5+{{16a^2}\over{v^2}}(c+2),\qquad
 \lambda_5={{4a^2}\over{v^2}}d,      \nonumber\\
 B_1&=& -{{\sqrt{2}a^2}\over v}e,\qquad
 B_2 =  -{{\sqrt{2}a^2}\over v}f,    \label{eq:dis-sym-para1}
\end{eqnarray}
and from the kink ansatz for the case of $\theta=0$,
the others are given by
\begin{eqnarray}
 m^2&=& 4a^2 - m_3^2,     \nonumber\\
 A+B_0&=& {{4\sqrt{2}a^2}\over v} - B_1 - B_2, \nonumber\\
 \lambda_6&=& {1\over4}(\tilde\lambda-\lambda_5)-{{8a^2}\over{v^2}}
           =  {{4a^2}\over{v^2}}c.       \label{eq:dis-sym-para2}
\end{eqnarray}
Once $(b,c,d,e,f)$, the wall thickness $1/a$ and the Higgs VEV
$v$ at the transition temperature are given, the effective potential
is determined.
Such an effective potential has the degenerate 
minima and covers a restricted class of the potentials of the form
(\ref{eq:dis-sym-Veff}). We expect that the solutions obtained below
will survive when we slightly deform the effective potential
from the restricted class.\par
\subsection{Energy density and BAU}
The nucleation rate of the bubble is roughly given by
$\exp(-4\pi R_C^2{\cal E}/T_C)$,
where $R_C$ is the radius of the critical bubble, which is 
approximated by $R_C\simeq\sqrt{3F_C/(4\pi av^2)}$.
Here $F_C$ is the free energy of the critical bubble and is found to
be about $(140\sim160)T$\cite{bubbleE}.
${\cal E}$ is the energy density of the bubble,
\begin{equation}
 {\cal E} = \int_{-\infty}^\infty dz \left\{
  \half\sum_{i=1,2}\left[\left({{d\rho_i}\over{dz}}\right)^2
    +\rho_i^2 \left({{d\theta_i}\over{dz}}\right)^2 \right]
   + V_{\rm eff}(\rho_1,\rho_2,\theta) \right\}.     \label{eq:energy-z}
\end{equation}
For ``the solution with kink ansatz'', this is given by
\begin{eqnarray}
 {\cal E} &=& {{av^2}\over3} \nonumber\\ 
 & &
 + av^2\sin^2\beta\cos^2\beta\int_0^1dy
 \left[ y(1-y)^3\left({{d\theta}\over{dy}}\right)^2
  +{{1-y}\over y} b\,[1-\cos(\theta+\delta)] \right.  \nonumber\\
 & &+{{(1-y)^2}\over y}\left( [c(1-y)-e](1-\cos\theta)
    \phantom{d\over 4}\right. \nonumber\\
 & &+\left.\left. [{d\over4}(1-y)-f][1-\cos(2\theta)] \right)
 \right]       \label{eq:kink-energy}       
\end{eqnarray}
where $av^2/3$ is the energy density of the trivial kink solution.
It was evaluated for several solutions in \cite{FKOTTb} and 
\cite{FKOTeCP}.
With the discrete symmetry, the energy density is reduced to
\begin{equation}
 {\cal E} =
 \int_0^1dy\left\{
  ay(1-y)\left[ \left({{d\rho(y)}\over{dy}}\right)^2 +
                {1\over4}\rho^2(y)\left({{d\theta(y)}\over{dy}}\right)^2
         \right]
 +{1\over{2ay(1-y)}}V_{\rm eff}(\rho,\theta) \right\}.
         \label{eq:dis-sym-energy} 
\end{equation}
\par
In the charge transport scenario\cite{NKC},  
the generated baryon number $n_B$ is 
proportional to the hypercharge flux $F_Y$, which is caused by the difference
of the reflection rates of chiral fermions by the $CP$-violating 
bubble wall. That is,
\begin{equation}
 {{n_B}\over s}\simeq
 {\cal N} {{100}\over{\pi^2g_*}}\cdot \kappa\alpha_W^4\cdot
 {{F_Y}\over{v_w T^3}}\cdot \tau T,
     \label{eq:BAU-flux}
\end{equation}
where $s$ is the entropy density, ${\cal N}$ a model-dependent constant
of $O(1)$ and $\kappa$ the coefficient of the sphaleron transition rate
in the symmetric phase given by $\kappa=1.09\pm0.04$\cite{Ambjorn}.
$\tau$ may be approximated by the mean free time or diffusion time
$D/v_w$ with the diffusion constant $D$ of the charge carrier, 
where $D\simeq1/T$ for quarks and $D\simeq(10^2\sim10^3)/T$ for 
leptons\cite{Joyce1}. 
It was shown that if the forward scattering is enhanced, even for the 
top quark, $\tau T\simeq10\sim10^3$ depending on $v_w$, where the 
maximum value is realized at $v_w\simeq1/\sqrt{3}$ \cite{NKC}. 
For this optimal case,
\begin{equation}
 {{n_B}\over s}\simeq 10^{-3} \cdot {{F_Y}\over{v_w T^3}}.
                      \label{eq:opt-BAU}
\end{equation}
Then the hypercharge flux $F_Y/(v_w T^3) \gtsim O(10^{-7})$ would be
necessary to explain the present BAU.\par
As we noted, if $CP$ is violated spontaneously, the bubble with
the profile $(\rho_i,\theta)$ degenerates with its $CP$-conjugate
partner $(\rho_i,-\theta)$. Then their nucleation probability and
dynamics will be completely the same so that the net baryon number
generated is completely canceled with each other.
Even if an explicit $CP$ violation will resolve the degeneracy,
two (or more) kinds of bubbles contribute to the baryon number in
opposite signs.
When we denote the baryon number generated by the $j$-th kind of 
bubble as $(n_B/s)_j$ and the nucleation rate of it as
$N_j=\exp(-4\pi R_C^2{\cal E}_j/T_C)$, the BAU may be approximately 
given by
\begin{equation}
 {{n_B}\over s} = {{\sum_j\left({{n_B}\over s}\right)_j N_j}\over
                   {\sum_j N_j}},    \label{eq:net-BA}
\end{equation}
where the expansion rate of each bubble is assumed to be almost the same.
We shall estimate it for some numerical solutions obtained
in the next section.
\section{Classification of Bubble Walls and Numerical\break Examples}
We show some solutions to the coupled equations (\ref{eq:sym-rho-z})
and (\ref{eq:sym-theta-z}), that is, ``the solutions without kink
ansatz''.
First, one must specify the boundary conditions in the broken
and symmetric phases.\par
By construction, $V_{\rm eff}(\rho,\theta)$ has
two degenerate minima.
In the broken phase, the boundary value $\theta_0$ in 
$(v,\theta_0)\equiv (\rho(y=0),\theta(y=0))$ is determined by minimizing
$V_{\rm eff}(v,\theta_0)$.
Namely, for a fixed $v$, \footnote{Strictly speaking, since the kink ansatz is
imposed when $\theta(y)\equiv0$, the effective potential satisfying
the ansatz has a minimum at a different point from the prescribed one
when $\theta\not=0$.}\  
$\theta_0$ is the solution to
\begin{equation}
 -m_3^2\sin(\theta_0+\delta)+
 \left[\half(\lambda_5\cos\theta_0+\lambda_6)v^2
       +\sqrt{2}v(B_1+4B_2\cos\theta_0)\right]\sin\theta_0=0.
          \label{eq:theta0-1}
\end{equation}
When the parameters are written in terms of $(b,c,d,e,f)$ by
(\ref{eq:dis-sym-para1}) and (\ref{eq:dis-sym-para2}), this is
expressed as
\begin{equation}
  b\sin(\theta_0+\delta) +
  \left( c + d\cos\theta_0 - e -4f\cos\theta_0\right)\sin\theta_0 =0.
          \label{eq:theta0-2}
\end{equation}
In general, for small $\absv{\delta}$, this has a solution
$\theta_0=O(\delta)$ (mod $\pi$). This corresponds to the
$CP$-conserving minimum when $\delta=0$. For $\delta=0$ and $d-4f<0$, if
$\absv{(b+c-e)/(d-4f)}<1$, there exists a nontrivial minimum
$\cos\theta_0=-(b+c-e)/(d-4f)$, which implies spontaneous $CP$ 
violation in the broken phase. Such a boundary value was considered in
our previous work\cite{FKOTTb} under the kink ansatz.
Even if $\theta_0$ for $\delta=0$ obtained above is very small,
for example, $O(10^{-3})$, 
a small $\delta$ happens to drive $\theta_0$ nonperturbatively to a
large value,
such as $O(10^{-1})$, as seen by solving (\ref{eq:theta0-2}).
If the parameters in the potential match this condition only at finite
temperature, this will be a new possibility to have sufficient $CP$
violation at the EWPT.\par
As for the symmetric phase, it is obvious that $\rho(y=1)=0$. 
Although $\theta_1\equiv\theta(y=1)$ seems to be undetermined,
it is not the case when $m_3^2\not=0$ ($b\not=0$).
This is because the equation for $\theta(y)$, (\ref{eq:sym-theta-z})
written in terms of $y$, has singularity at $y=1$, which makes
the solution singular there unless $\sin(\theta_1+\delta)$ multiplying
the most singular $m_3^2$-term vanishes.
For such a singular solution the energy density diverges so that it cannot be
realized dynamically.\footnote{In practice, the integration region in
(\ref{eq:energy-z}) is finite so that the energy density is finite.
We, however, expect such a solution is hard to be nucleated because of
its large energy.}\  
Hence the boundary condition for $\theta(y)$ in the symmetric phase is
\begin{equation}
   \theta_1+\delta = n\pi,\qquad (n\in{\bf Z}). \label{eq:theta1}
\end{equation}
\par
As long as the effective potential is parameterized as in the previous 
section and $\absv{\delta}$ is sufficiently small 
$(\absv{\delta}<10^{-3})$, the possible lowest energy solutions will be
classified in terms of the boundary values of $\theta(y)$
as in the table below.
\begin{center}
 \begin{tabular}{c||c|c}
 \hline
    & $\delta=0$    &   $\delta\not=0$   \\
 \hline
 (A)& $\theta(y)\equiv0$ (trivial solution) &
      $\theta(y)=O(\delta)$ for ${}^\forall y$ \\
 \hline
 \lw{(B)}
    & $\theta_0=\theta_1=0$ & $\theta_0=O(\delta)$ and $\theta_1=-\delta$ \\
    & spontaneous $CP$ violation in the wall 
    & $\absv{\theta(y)}\ggt\absv{\delta}$ at $y\sim1/2$ \\
 \hline
 \lw{(C)}
    & $\theta_0\not=0$ and $\theta_1=0$ 
    & $\absv{\theta_0}\ggt O(\delta)$ and $\theta_1=-\delta$ \\
    & spontaneous $CP$ violation in the broken phase 
    & $\absv{\theta(y)}\ggt\absv{\delta}$ at $y\ltsim 1/2$ \\
 \hline
 \lw{(D)}
    & $\theta_0=0$ and  $\theta_1=\pi$ 
    & $\theta_0=O(\delta)$ and $\theta_1=\pi-\delta$ \\
    & maximal $CP$ violation in the wall 
    & maximal $CP$ violation in the wall \\
 \hline
 \end{tabular}
\end{center}
\par
As an example of type (A) with $\delta\not=0$, we study the case 
for $(b,c,d,e,f)=(3,5,5,7,$
$-1.25)$.
We plot the wall profile or the mass function in the Dirac equation,
$(\rho\sin(\theta/2),$
$\rho\cos(\theta/2))$, in Fig.~1 and
the chiral charge flux obtained from it in the dimensionless
unit of $F_Q/(v_wT^3$
$(Q_L-Q_R))$ in Fig.~2, where the wall velocity
and temperature are taken to be 
$v_w=1/\sqrt{3}$ and $T=100\mbox{GeV}$, respectively.
$Q$ denotes a certain quantum number such as the hypercharge.
%
\begin{figure}
 \epsfxsize = 8cm
 \centerline{\epsfbox{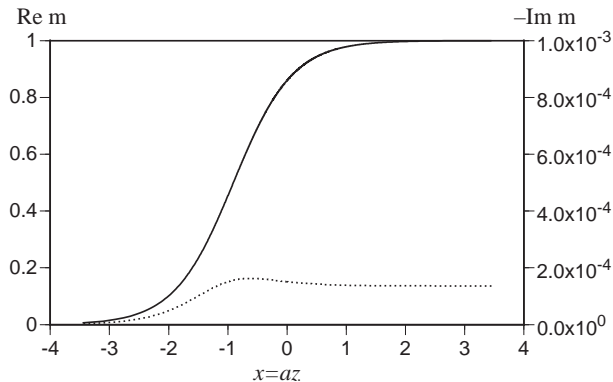}}
 \caption{The wall profile for $(b,c,d,e,f)=(3,5,5,7,-1.25)$ and 
 $\delta=10^{-3}$ as a function of the dimensionless variable $x=az$.
 Note that the $CP$-violating imaginary part ( dotted curve ) is much
 smaller than the real part ( solid curve ).}
 \label{fig:1}
\end{figure}
%
\begin{figure}
 \epsfxsize = 8cm
 \centerline{\epsfbox{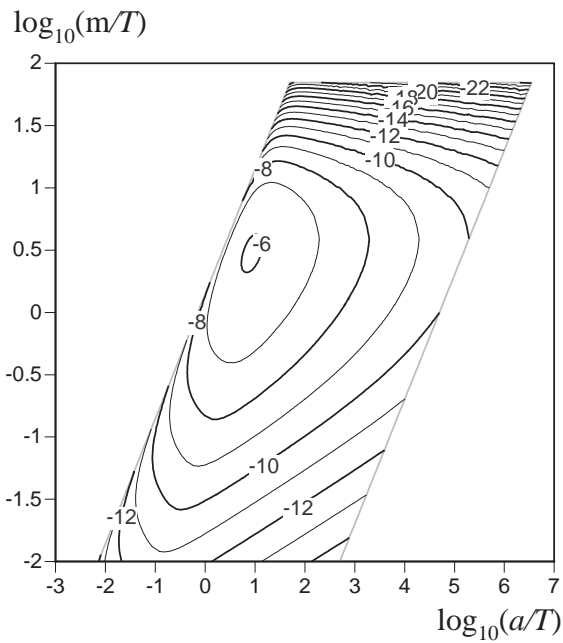}}
 \caption{Contour plot of the net chiral charge flux, normalized as
 $\log_{10}\left[-F_Q/(v_w T^3(Q_L-Q_R))\right]$ for the profile shown
 in Fig.~1. Here we take $v_w=0.58$ and $T=100\mbox{GeV}$.}
 \label{fig:2}
\end{figure}
For this type of solution with $\theta(y)=O(\delta)$,
it seems difficult to have sufficient BAU except for
a restricted set of the carrier mass and the wall thickness.\par
If a solution of type (D) exists, it will incorporate so large $CP$ 
violation near the bubble wall to generate sufficient BAU.
We have attempted to find such a solution by adjusting the parameters
in the effective potential, but have not succeeded.
This may be because the cost of the derivative term in the energy
cannot be compensated by the gain of the potential term along such
a path from the broken to the symmetric phase.\par
If type (D) solution does not exist, the only possible one
accompanying large $CP$ violation $\theta\sim O(1)$
in an intermediate range will be of the 
type (B) or (C).
The range where $\theta$ is large should not be near the symmetric phase,
where $\rho(y)\simeq0$, since the $CP$ violation in the Yukawa interactions
is proportional to $\rho(y)\sin(\theta(y)/2)$.
A solution of type (B) for $\delta\not=0$ was found under the the kink 
ansatz\cite{FKOTeCP}.\par
Let us denote the solutions of the $CP$-conjugate pair in the
case of $\delta=0$ as $(\rho,\theta)$ and $(\rho,-\theta)$.
For sufficiently small $|\delta|\not=0$, there generally appear three
solutions, $(\rho^+,\theta^+)$, $(\rho^-, \theta^-)$ and
$(\rho^0, \theta^0)$, such that they respectively approach
$(\rho,\theta)$, $(\rho,-\theta)$ and the trivial kink solution
$(\rho_{\rm kink},\theta=0)$ as $\delta\rightarrow 0$.
Note that $(\rho^0,\theta^0)$ has no $CP$-conjugate partner.
An example presented here employs the same parameter set and the 
boundary condition as in \cite{FKOTeCP}
to solve the equations (\ref{eq:sym-rho-z}) and
(\ref{eq:sym-theta-z}) for various $\delta$ starting from $\delta=0$.
That is, $(b,c,d,e,f)=(3,12.2,-2,12.2,0)$
and $\theta_1=-\delta$ while $\theta_0$ is determined by
(\ref{eq:theta0-2}).
There are three numerical solutions, $(\rho^0,\theta^0)$, 
$(\rho^+,\theta^+)$ and $(\rho^-,\theta^-)$, in the order of the 
decreasing energy, satisfying the same boundary conditions for 
$\delta\ltsim 0.0045$. For larger $\delta$, only $(\rho^-,\theta^-)$ exists.
Their energy densities, measured from that of the trivial kink configuration,
$\delta{\cal E}\equiv{\cal E}- av^2/3$, are plotted in Fig.~3 in the unit of $av^2$.
%
\begin{figure}
 \epsfxsize = 8cm
 \centerline{\epsfbox{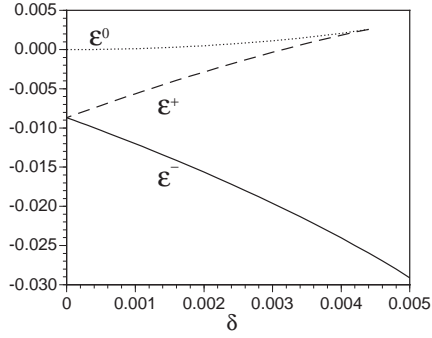}}
 \caption{Energy densities of the bubble walls for 
 $(b,c,d,e,f)=(3,12.2,-2,12.2,0)$ measured from the trivial kink
 solution in the unit of $av^2$.}
 \label{fig:3}
\end{figure}
To illustrate how $CP$ is violated in the intermediate region, we
show the contour plot of the effective potential together with the 
path $(\rho^+,\theta^+)$ for $\delta=10^{-3}$ in Fig.~4. 
%
\begin{figure}
 \epsfxsize = 9cm
 \centerline{\epsfbox{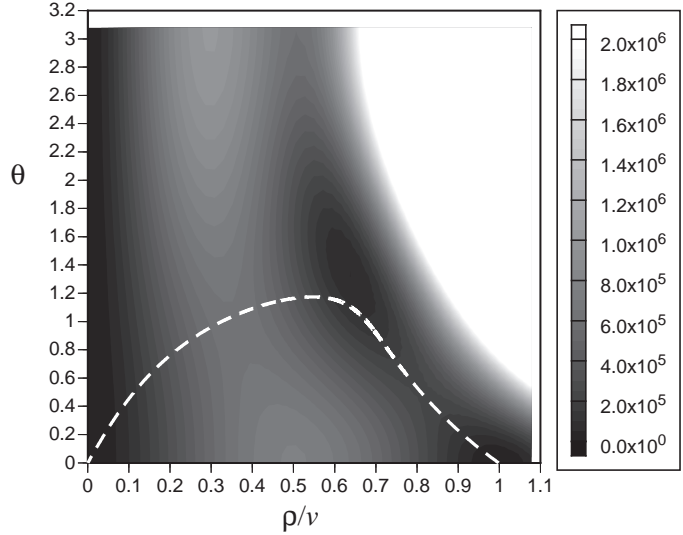}}
 \caption{Contour plot of the effective potential together with the
 path of $(\rho^+,\theta^+)$ connecting the two vacua for the same parameter
 set as Fig.~3 and
 $\delta=10^{-3}$.}
 \label{fig:4}
\end{figure}
The mass functions of each profile for $\delta=10^{-3}$ are plotted in Fig.~5. 
%
\begin{figure}
 \epsfxsize = 10cm
 \centerline{\epsfbox{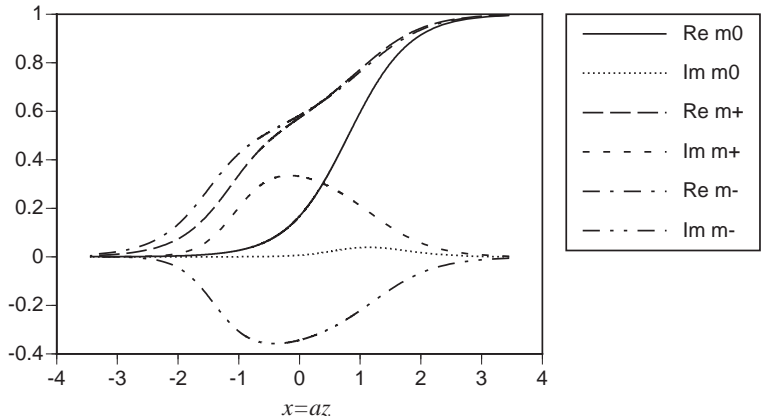}}
 \caption{The wall profiles or the mass functions for
 $(b,c,d,e,f)=(3,12.2,-2,12.2,0)$ and 
 $\delta=10^{-3}$ as functions of $x=az$.}
 \label{fig:5}
\end{figure}
For illustration to evaluate BAU, we
take the case of $\delta=10^{-3}$.
Since the energy densities of them are, 
\begin{eqnarray}
 \delta{\cal E}[\rho^0,\theta^0]&=& 1.21567\times10^{-4}av^2,\nonumber\\
 \delta{\cal E}[\rho^+,\theta^+]&=&-5.6350\times10^{-3}av^2,
                                                 \label{eq:energy-ex1}\\
 \delta{\cal E}[\rho^-,\theta^-]&=&-1.2012\times10^{-2}av^2, \nonumber
\end{eqnarray}
the ratios of the nucleation rates are
\begin{equation}
 {{N_+}\over{N_0}} = 12.23, \qquad\qquad
 {{N_-}\over{N_0}} = 196.0,  \label{eq:enhance-ex1}
\end{equation}
where we have used the expression for the nucleation rate of the 
critical bubble
\begin{equation}
 N\simeq \exp\left(-{{4\pi R_C^2{\cal E}}\over{T_C}}\right),
       \label{eq:nucleation-rate}
\end{equation}
with $R_C\simeq\sqrt{3F_C/(4\pi av^2)}$ and $F_C\simeq145T_C$.
By neglecting an extremely small contribution from
$(\rho^0,\theta^0)$, as understood from the small imaginary part
in Fig.~5, it is found that about 88 \% of baryon number generated
by $(\rho^-,\theta^-)$ survives the cancellation by
$(\rho^+,\theta^+)$.
Thus $\delta=10^{-3}$ will be sufficient to leave the BAU for
wide range of the carrier mass and the wall thickness.
As seen from Fig.~3, only one species of the bubbles exist for
$\delta>0.0045$ so that the cancellation among the bubbles does
not occur.
The net chiral charge flux (\ref{eq:net-BA}) for $\delta=10^{-3}$ 
is presented in Fig.~6.
%
\begin{figure}
 \epsfxsize = 8cm
 \centerline{\epsfbox{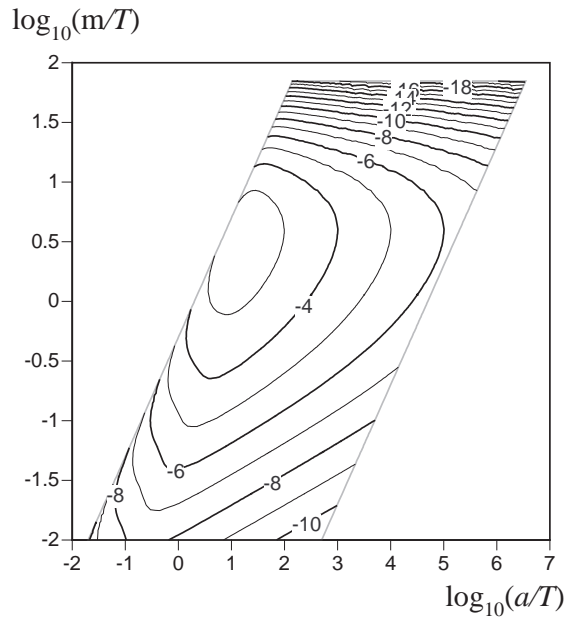}}
 \caption{Contour plot of the net chiral charge flux, normalized as
 $\log_{10}\left[-F_Q/(v_w T^3(Q_L-Q_R))\right]$ for the profile shown
 in Fig.~5. Here we take $v_w=0.58$ and $T=100\mbox{GeV}$.}
 \label{fig:6}
\end{figure}
\par
Some solutions of type (C) for $\delta=0$ were found under the the kink 
ansatz\cite{FKOTTb}.
The profile for non-vanishing $\delta$ is similar to those presented
there\cite{FKOTTb}, which monotonously connects $(v,\theta_0)$ to
$(0,\theta_1=-\delta)$. The magnitude of the chiral charge flux is 
controlled by that of $\theta_0$.
\section{Conclusion and Discussions}
In conclusion, assuming that the EWPT is of first order accompanying 
nucleation and constant growth of the broken-phase bubbles within the
symmetric phase and that the dynamics of the bubbles are governed by
the classical equations of motion of the gauge-Higgs sector, we have
given some examples of
the bubble profile by solving the equations of motion derived from
the effective potential.
According to the boundary conditions, the possible types of solutions are 
classified.
Among them, the profiles of type (B), (C) and (D) will yield large $CP$
violation around the bubble wall, though the type (D) seems hard to be
realized dynamically. The solutions of type (B) require
the parameters to admit spontaneous $CP$ violation in some region
between the broken and symmetric phases.
The explicit $CP$ violation, which is small enough not in contradiction with
observations, can avoid the cancellation of the generated baryon number
between the $CP$-conjugate bubbles, leading to the BAU. 
The type (A) bubbles will yield insufficient BAU except for a 
restricted set of the carrier mass and the wall thickness.\par
If the temperature-dependent parameters are calculated in the MSSM at the 
one-loop level,
the parameter $f$ will be generated only by higher order 
terms in the high-temperature expansion of the squark or
slepton contribution\cite{FKOTc}, and
the condition $d\rho^2-4f\rho<0$ will be satisfied by $d<0$.
For $d=v^2/(4a^2)\cdot\lambda_5<0$ to be realized, we must have
tree-level negative $\lambda_5$ in the two-Higgs-doublet model or
radiatively induced negative $\lambda_5$ in the MSSM.
In this case, we must avoid the problems peculiar to the spontaneous 
$CP$ violation, that is, too light scalar and the domain wall.
Both will be cured by a small explicit $CP$ violation.
In the MSSM, it will be unnatural to assume that all the soft 
supersymmetry breaking parameters do not introduce any explicit $CP$ 
violation at the electroweak scale.
As we noted in the previous section, if $CP$ is spontaneously violated
in the broken phase when $\delta=0$, a small $\delta$ 
nonperturbatively happens to drive large $CP$ violation.
This will further restrict some set of the parameters in the model.
These constraints will be examined in a future publication.\par
We have presented the chiral charge flux for the profile of type (A) 
and (B), which is calculated supposing that one of the Higgs doublets
couple to a fermion mass eigenstate.
In practice, the weak hypercharge plays its role and
the hypercharge flux is {\it additive} for each fermion species in the
standard model.
This is because the up-type quark with positive hypercharge couples to
$e^{i\theta_1}=e^{i\theta/2}$, while the down-type quarks or the 
leptons with negative hypercharge couples to $e^{i\theta_2}=e^{-i\theta/2}$
or $(e^{i\theta_1})^*=e^{-i\theta/2}$
in the MSSM or the two-Higgs-doublet models with the 
discrete symmetry to avoid the tree-level FCNC.
In a supersymmetric standard model, there also exist contributions 
from the superpartners with nonzero hypercharge such as the scalar 
partners of the fermions and those of the Higgs scalars. The effects 
of the chargino were considered in \cite{Aoki}.\par
Finally, we comment on the discrete symmetry (\ref{eq:def-discrete-sym}),
which is a working hypothesis to reduce the dynamical degrees of freedom.
The numerical method we employed to solve the boundary value problem
is the relaxation algorithm, which varies the configuration to approach
the correct solution starting from a prescribed initial configuration.
Toward our final goal to find solutions to the equations for the
full dynamical degrees without the discrete symmetry,
the knowledge of the solutions obtained here will help
us to prepare an appropriate initial condition for these equations.
That is, as long as the solution with the discrete symmetry is stable
in the enlarged functional space, we can obtain the desired solution
by graduately modifying the parameters from those with the discrete
symmetry to find the next initial configuration, step by step.
In fact, we checked that for the solution of type (B), the discrete 
symmetry is not broken spontaneously, that is, the solution with
the discrete symmetry is still a solution to the equations of 
motion for all possible order parameters.
%
%

%
\end{document}